\documentclass[preprint]{aastex}
\usepackage{epsfig}

\shortauthors{Mattson \& Weaver}
\shorttitle{X-Ray Spectral Variability of MCG $-$5-23-16}

\begin{document}

\newcommand{\RXTE}{\textit{RXTE}}
\newcommand{\ASCA}{\textit{ASCA}}
\newcommand{\SAX}{\textit{BeppoSAX}}
\newcommand{\nH}{N$_{\rm H}$}

\title{\RXTE\ and \SAX\ Observations of MCG $-$5-23-16:
Reflection From Distant Cold Material}

\author{B. J. Mattson\altaffilmark{1} and K. A. Weaver}
\affil{NASA/Goddard Space Flight Center, Laboratory for High Energy 
Astrophysics, Greenbelt, MD 20771}

\altaffiltext{1}{Also L-3 Communications Government Services, Inc.}

\begin{abstract}
We examine the spectral variability of the Seyfert 1.9 galaxy
MCG$-$5-23-16 using \RXTE\ and \SAX\ observations spanning 2 years from
April 1996 to April 1998.  During the first year the X-ray source
brightens by a factor of $\sim25\%$ on timescales of days to months. 
During this time, the reprocessed continuum emission seen with \RXTE\
does not respond measurably to the continuum increase.  However, by the
end of the second year during the \SAX\ epoch the X-ray source has faded
again.  This time, the reprocessed emission has also faded, indicating
that the reprocessed flux has responded to the continuum.  If these
effects are caused by time delays due to the distance between the X-ray
source and the reprocessing region, we derive a light crossing time of
between $\sim1$ light day and $\sim1.5$ light years. This corresponds to
a distance of 0.001 pc to 0.55 pc, which implies that the reprocessed
emission originates between $3\times10^{15}$ cm and $1.6\times10^{18}$
cm from the X-ray source.  In other words, the reprocessing in
MCG$-$5-23-16 is {\it not} dominated by the inner regions of a standard
accretion disk.
\end{abstract}

\section{Introduction} \label{section:intro}

X-ray variability studies can probe the geometry of active galactic
nuclei by giving clues about where X-ray reprocessing occurs in relation
to the central X-ray source.  According to current unified models for
Seyfert 1 and Seyfert 2 galaxies (Antonucci 1993), two likely locations
for reprocessing are an accretion disk (George \& Fabian 1991; Matt,
Perola \& Piro 1991) and an obscuring molecular torus (Ghisellini,
Haardt \& Matt 1994; Krolik, Madau \& \.{Z}ycki 1994). The X-ray
spectral features of reprocessing consist of an Fe K$\alpha$
fluorescence line at $\sim6.4$ keV and a ``hump'' due to Compton
reflection, which starts to dominate at $\sim10$ keV and is produced by
the combined effects of photoelectric absorption and Compton
downscattering. If reprocessing occurs in the inner regions of an
accretion disk, we expect to see Doppler and relativistically broadened
spectral features and small time lags (minutes to hours) between changes
in the intrinsic X-ray flux and changes in the iron and reflection
components. However, if reprocessing occurs in a larger region such as a
molecular torus, then the reflected flux is controlled by the
time-averaged primary spectrum rather than the instantaneous (observed)
one (Malzac and Petrucci 2001), and we might observe a substantial time
lag (on the order of years) between changes in the continuum flux and
changes in the reflection features.

\begin{figure}[h]
  \epsfxsize=3in
  \rotatebox{270}{\epsfbox[0 0 560 620]{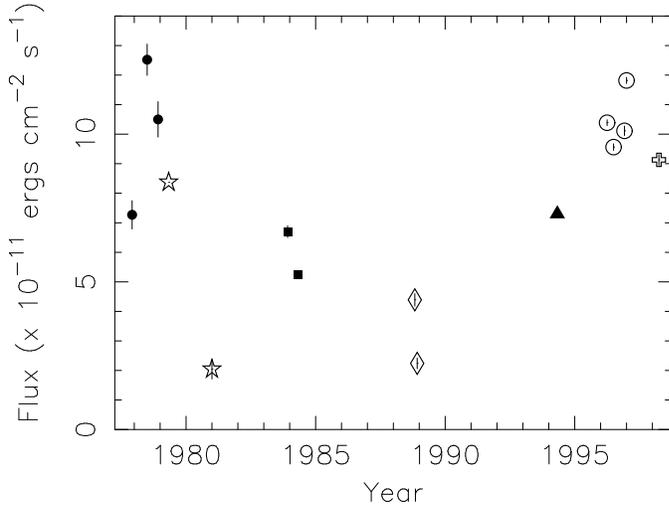}}
  \caption{The long-term 2-10 keV flux 
variability of MCG $-$5-23-16.  The symbols represent 
data from different satellites: filled circles from HEAO1 (Mushotzky 1982), 
stars from {\it Einstein} (Turner et al. 1991; Kruper, Canizares \& 
Urry 1990), squares from EXOSAT (Singh, Rao \& Vahia 1992), diamonds from 
{\it Ginga} (Nandra \& Pounds 1994), triangle from \ASCA\ 
(Weaver et al. 1997), open circles from \RXTE\ (this paper), and cross 
from \SAX\ (this paper).  A second \ASCA\ observation overlaps the third 
\RXTE\ observation.}  
  \label{fig:longterm}
\end{figure}

MCG $-$5-23-16 (z = 0.008) is an X-ray bright AGN and optically
classified as a Seyfert 1.9 galaxy (V\'{e}ron et al. 1980). The 2 to 10
keV flux varies by at least a factor of 6 between a low state of $\sim 2
\times 10^{-11}$ ergs cm$^{-2}$ s$^{-1}$ and a high state of $\ga 12
\times 10^{-11}$ ergs cm$^{-2}$ s$^{-1}$, as shown in
Figure~\ref{fig:longterm}. An \ASCA\ observation shows a complex,
triple-peaked Fe K$\alpha$ line profile (Weaver et al. 1997), which is
attributed to a combination of a double-peaked emission line produced in
the inner regions of an accretion disk and a narrow emission line from
further out in the disk or from elsewhere in the galaxy. The key to the
origin of the narrow component is its variability. Based on its line
width, we might assume that it originates from a distance that is at
least $10^3 R_s$ ($R_s = 2GM/c^2 =$ Schwarzchild radius) from the black
hole, which translates into a light crossing time for a $10^8$M$_\sun$
black hole of $t \sim 10^5$ s, or a few days.

We present the first broad band spectral variability study of MCG
$-$5-23-16, using data from the \textit{Rossi X-Ray Timing Explorer}
(\RXTE) and \SAX.  A previous analysis of one of the \RXTE\ observations
revealed the signature of Compton reflection and confirmed the presence
of the broad Fe K line (Weaver, Krolik \& Pier 1998).  Here we begin
with the knowledge that these features exist and we study their
variability behavior.

\section{Data Analysis} \label{section:analysis}

\subsection{\RXTE\ Observations} \label{subsection:xte_model}

\begin{figure}[p]
  \epsfxsize=3in
  \rotatebox{270}{\epsfbox[0 0 560 620]{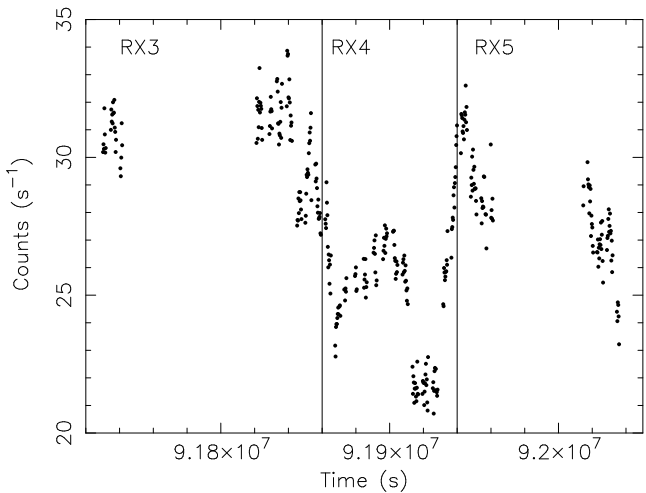}}
  \caption{\RXTE\ PCA lightcurve of the 2 to 10 keV flux for the 
100 ks long-look.  The vertical lines indicate the temporal divisions 
that were used for extracting spectra.  Mean fluxes for each 
temporal bin are plotted in Figure~\ref{fig:midterm}.}
  \label{fig:shortterm}
\end{figure}

\begin{figure}[p]
  \epsfxsize=3in
  \rotatebox{270}{\epsfbox[0 0 560 620]{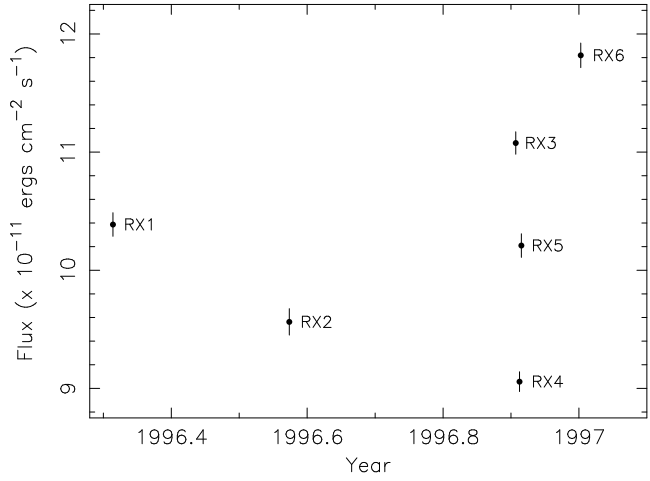}}
  \caption{Time-averaged lightcurve for the six \RXTE\ spectra.}
  \label{fig:midterm}
\end{figure}

Four observations of MCG $-$5-23-16 were made by the \RXTE\ Proportional
Counter Array (PCA): three short 25 ks observations 24 April 1996, 28-29
July 1996 and 10 January 1997, and one long 100 ks observation 27-30
November 1996. The PCA consists of five Proportional Counter Units
(PCUs) and is sensitive to energies from 2 to 60 keV. The Standard 2
data for each observation were reduced with the REX (v0.2) script for
Ftools (v5.0.1), using the default selection criteria
(http://heasarc.gsfc.nasa.gov/docs/xte/recipes/rex.html). Those criteria
are: data from layer 1 and from PCUs 0, 1 and 2; times with Earth
elevation angle greater than 10$^\circ$, pointing offset less than
0.02$^\circ$ and electron contamination less than 0.1; and times greater
than 30 minutes after the last passage through the South Atlantic
Anomaly (SAA). Based on the light curve in Figure~\ref{fig:shortterm},
the long observation was split into two high states and a low state and
spectra of each were extracted separately. This provides six spectra,
which we refer to as RX1 (April 1996), RX2 (July 1996), RX3, RX4, and
RX5 (November 1996), and RX6 (January 1997). Figure~\ref{fig:midterm}
shows the time averaged lightcurve for the six \RXTE\ spectra.

The background was estimated using PCABACKEST (v2.1e). We used the
L7-240 background model which incorporates two background components.
The ``L7'' background estimates the internal detector background and is
based on housekeeping parameters read directly from the Standard 2 data
files. The ``240'' background approximates the effects of particles present
in the detectors after each SAA passage by modeling them  
as a decaying exponential with a half-life of 240 s 
(http://heasarc.gsfc.nasa.gov/docs/xte/pca\_news.html). The response
matrix was generated using the Ftool PCARSP (v2.43) with the
channel-to-energy matrix from the \RXTE\  Guest Observer Facility (GOF)
web site (http://heasarc.gsfc.nasa.gov/docs/xte/xte\_1st.html).
We are confident in
the instrument response and background models up to energies of $\sim$25
keV.  The background begins to dominate above 25 keV and so we ignore
channels with higher energies.

About 11 ks of the long-look 100 ks observation had previously been analyzed by
Weaver, Krolik and Pier (1998), who found less Compton reflection in 
the spectrum of MCG$-$5-23-16 than our analysis indicates.  
Since then, the Q6 background models used by Weaver et al.\ have been retired
(http://heasarc.gsfc.nasa.gov/docs/xte/pca\_news.html) and the response matrices
have changed to reflect the time-dependency of a drift in the detector
gain and an increasing xenon fraction in the propane layer of the PCA
detectors. In addition, new 
background models, which refine the L7-240 models, have been made
available since our analysis of MCG$-$5-23-16 was completed.  We
have examined other data similar to MCG$-$5-23-16 using both the 
new and the old L7-240
models and find that the results are consistent.  We therefore do not 
expect the new background models to alter our results. 
Note however that we do not make specific claims to have reliably
measured the absolute amount of reflection in MCG$-$5-23-16, only 
changes in the {\it relative} amount of reflection.

\subsection{\SAX\ Observation} \label{subsection:SAX_model}

\SAX\ performed a 100 ks observation of MCG$-$5-23-16 on 24 April 1998.
The  Low Energy Concentrator Spectrometer (LECS) and Medium Energy
Concentrator Spectrometers (MECS) are imaging telescopes sensitive to
0.1 to 10 keV (LECS) and 1.3 to 10 keV (MECS); whereas, the collimated
Phoswich Detector System (PDS) is sensitive to energies from 15 to 300
keV. We only use data from the MECS2 and MECS3, added together. On May
6, 1997 the MECS1 unit developed a fault in its gas cell high voltage
supply and has been unavailable since.

We generated images from the LECS and the combined MECS2 and MECS3 data
using XSELECT version 2.0 and the event files available through the
\SAX\ Science Data Center's web archive (http://www.sdc.asi.it/). The
source and background spectra were then extracted based on these images.
We obtained a background-subtracted PDS spectra directly from the Data
Center's web archive. The source flux remained relatively constant
during the \SAX\ observation, so we extract the total spectrum. The
response matrices and arf files were obtained from the \SAX\  Guest
Observer U.S. Coordination website at GSFC
(ftp://heasarc.gsfc.nasa.gov/sax/cal/responses/98\_11). We adopt the
recommended energy ranges of  0.12 to 4.0 keV for the LECS, 1.65 to 10.5
keV for the MECS, and 15.0 to 220.0 keV for the PDS.  We use the full
energy band because, although MCG$-$5-23-16 is heavily absorbed, there
are soft counts in the Rosat band (Mulchaey et al. 1993).

\section{Results of Spectral Fits} \label{section:results}

\subsection{\RXTE\ Spectra} \label{subsection:xte_results}

The spectrum from each \RXTE\ epoch (RX1 to RX6) is fitted from 2.5 to
25 keV with an absorbed Compton reflection model (PEXRAV in XSPEC) plus
a Gaussian. PEXRAV simulates the effects of an exponentially cut-off
power-law reflected by neutral (except H and He) matter (Magdziarz \&
Zdziarski 1995). There are seven model parameters, which are the photon
index of the intrinsic, underlying power-law ($\Gamma$), the cutoff
energy of the power law in keV ($E_c$), the relative amount of
reflection ($R$),  the redshift ($z$), the abundance of heavy elements
in solar units ($Z$), the disk inclination angle ($i$), and the photon
flux of the power law at 1 keV in the observer's frame ($A$). The
relative amount of reflection is normalized to 1 for the case of an
isotropic source above a disk of neutral material ($\Omega=2\pi$).
Adding a Gaussian line (energy in keV, physical width ($\sigma$) in keV
and flux in units of photons cm$^{-2}$ s$^{-1}$) and an
absorbing column (\nH\ in units of cm$^{-2}$) yields a total of 12 parameters.

We fix the following values in PEXRAV: $E_c$ = 100 keV, $z$ = 0.008, $Z$
= 1.0 and $\cos i$ = 0.95. The cutoff energy is frozen at 100 keV for
simplicity since we obtain fairly large errors on this value when
fitting the \SAX\ data ($E_c\sim180+/-100$).  Perola et al. (2002) find
a more conclusive value of $E_c = 147^{+70}_{-40}$.  But in any case,
fixing the cutoff energy at 147 does not affect our results. The
inclination angle of the accretion disk is assumed to be face-on to
allow comparison of our results with those of other authors. We also
hold the Gaussian width at its best-fitting mean value of $\sigma$ =
0.24 keV. This leaves six free parameters: $\Gamma$, $R$, $A$, iron line
energy, iron line flux and \nH.

\begin{figure}[p]
  \epsfxsize=3in
  \rotatebox{270}{\epsfbox[0 0 560 620]{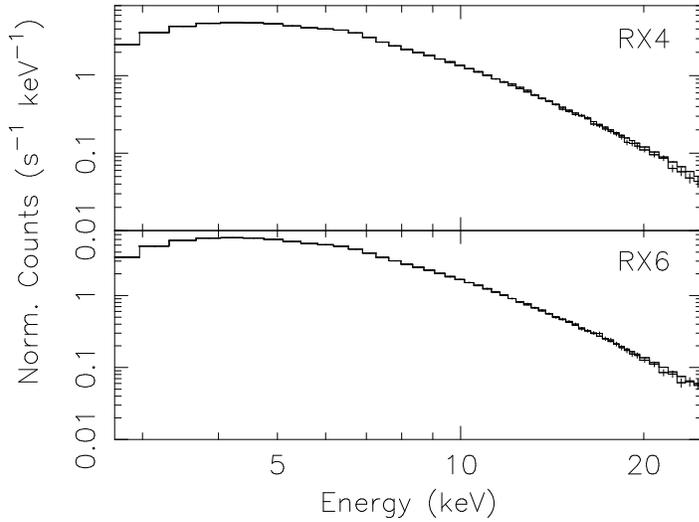}}
  \caption{\RXTE\ spectra for the lowest (RX4) and highest (RX6) flux states.}
  \label{fig:spec_rxte}
\end{figure}

The initial spectral results are listed in Table~\ref{table:free_gamma}. 
All error bars are 90\% confidence for one interesting parameter.
Naively fitting the spectra with all parameters
free provides the following results: $\Gamma$ ranges from $1.85^{+0.03}_{-0.03}$ (RX2)
to $2.00^{+0.02}_{-0.01}$ (RX6), \nH\ remains approximately constant
at $2.9\times10^{22}$ cm$^{-2}$ 
($\chi^2/\nu$ = 0.88/6 for a constant value) and $R$ changes from
$\sim0.4$ (RX1) to $\sim0.8$ (RX6).
The first five \RXTE\ spectra (RX1 through RX5) are similar in shape
while RX6 is remarkably softer with a larger apparent amount of reflection 
and an apparently weaker Fe K line (EW of $\sim113$ eV compared 
to $\sim125-144$ eV).  
For comparison we show the spectra from
the highest and lowest flux states (RX4 and RX6) in
Figure~\ref{fig:spec_rxte}. The cause of the spectral softening in RX6 is not clear at
first glance because there are model degeneracies between $\Gamma$, \nH\,
and $R$ that can occur as a result of the
way these parameters trade off against each other in the
modeling process.  

The large modeling degeneracies between the $\Gamma$, \nH\, and $R$
can lead to false conclusions about spectral variability and so we 
are challenged to deal with them in a way that allows us to eliminate the 
degeneracies as much as possible and focus on what we believe is the
true variability in the source.
But how can we know which parameters represent real changes in the X-ray
source and which are artifacts of the modeling?  For example, $R$ has increased
substantially in RX6 and this increase is consistent with a
systematic error that may be caused by a degeneracy in the model.
This means that we need to know if there is a real change in the 
continuum photon index of the source.  
Unfortunately, when we attempt to measure $\Gamma$ 
by using only the data between 2.5 and 5.5 keV, 
to avoid contamination by reflection, the errors ($\sim \pm
0.13$) are too large to be useful.

\begin{figure}[p]
  \epsfxsize=3in
  \rotatebox{270}{\epsfbox[0 0 560 620]{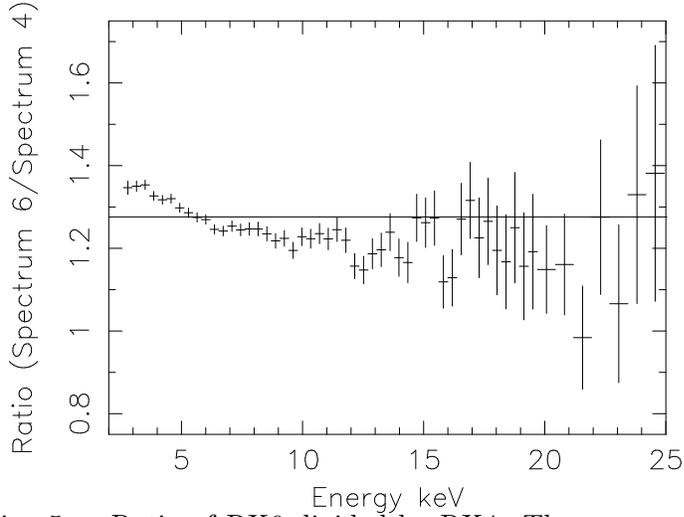}}
  \caption{Ratio of RX6 divided by RX4.  These 
correspond to the highest and lowest flux states observed with \RXTE.  The
straight line represents the best fitting constant to this ratio ($\chi^2/\nu = 385.5/65$).
This plot illustrates the spectral softening below 5 keV that we believe is  
caused by a decrease in the absorbing
column density between the two observations.}
  \label{fig:ratio}
\end{figure}

To eliminate any systematic errors that have been introduced by the modeling,  
the instrumental response or the background estimations, 
we calculate the ratio of RX6 to RX4
(Figure~\ref{fig:ratio}).  The spectral variability is apparent mainly at
energies $\la 5.5$ keV. An absorbing column density of $3\times10^{22}$
cm$^{-2}$ depletes the spectrum at energies up to around 5 keV and 
so Figure~\ref{fig:ratio} implies that the absorbing column density,
rather than $\Gamma$, has changed during the high state.  Since 
the energy bandpass of the PCA is limited, the constraints 
on \nH\ are actually worse than the statistical errors would indicate.
The modeling procedure compensates for this spectral change by
varying (increasing) $\Gamma$ and $R$ to achieve a good statistical fit
because of the mathematical
degeneracy in the reflection model.\footnote{This three-way degeneracy
is especially a problem for MCG$-$5-23-16 because it possesses just
enough X-ray absorption, a few times $10^{22}$ cm$^{-2}$, to cut the
spectrum off near the lower end of the \RXTE\ bandpass, which confuses
the PCA due to its relatively poor energy resolution.  Whether this degeneracy is as
severe for other bright Seyfert galaxies has not been tested.}

\begin{figure}[p]
  \epsfxsize=4in
  \rotatebox{0}{\epsfbox[0 0 589 487]{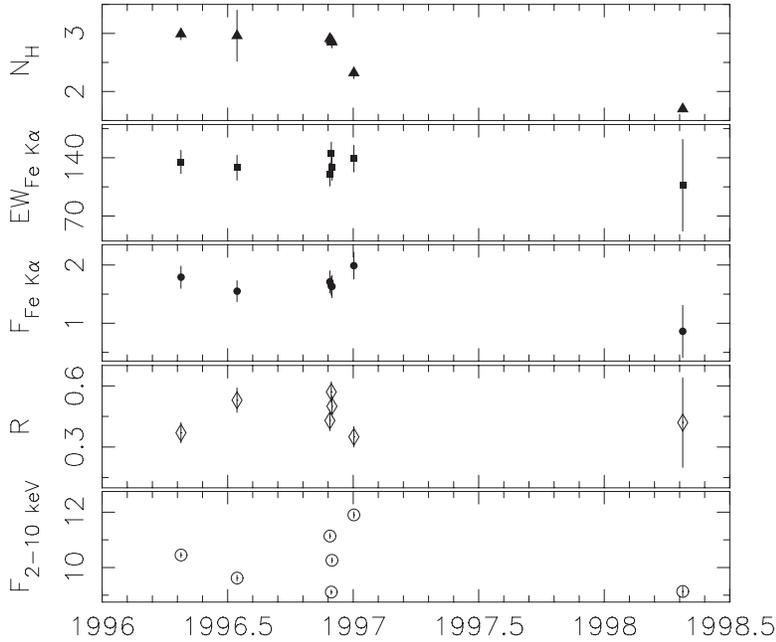}}
  \caption{Model parameters versus time.  From top to bottom are the absorbing 
column in units of $10^{22}$ cm$^{-2}$, Fe K equivalent width in eV,
Fe K line flux in units of $10^{-4}$ photons cm$^{-2}$ s$^{-1}$,
reflection fraction ($R$) and $2-10$ keV continuum flux in 
units of $10^{-11}$ erg cm$^{-2}$ s$^{-1}$, respectively.}
  \label{fig:stack}
\end{figure}

\begin{figure}[p]
  \epsfxsize=3in
  \rotatebox{270}{\epsfbox[0 0 560 620]{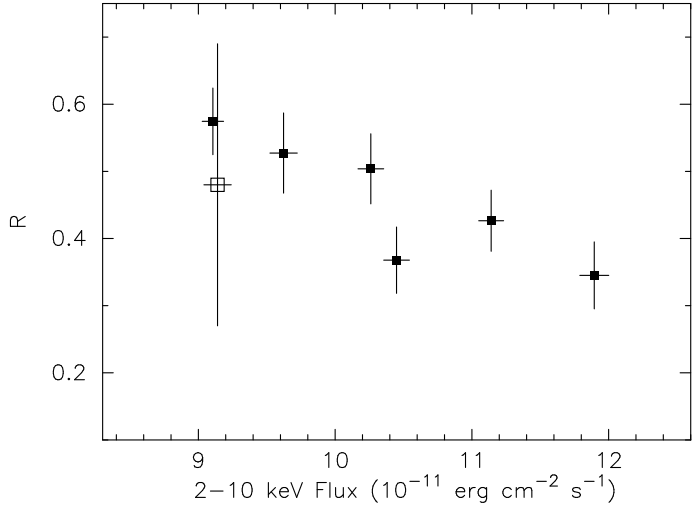}}
  \caption{The reflection fraction versus 2 to 10 keV 
continuum flux.  The solid squares are the \RXTE\ points and the open 
square is the \SAX\ point.}  
  \label{fig:fvr}
\end{figure}

We conclude that a decrease in absorption, rather than a change in the 
photon index, dominates the spectral variability in RX6.
We therefore feel confident assuming that $\Gamma$ does not change
significantly during the \RXTE\ epoch and adopt $\Gamma = 1.88$ (the mean value for
RX1 through RX5) for the remaining spectral fits.
Table~\ref{table:fixed_gamma} lists the new results for the Compton 
reflection model, this time 
fixing $\Gamma$ and leaving \nH\ and $R$ free.  The model parameters are
plotted in Figure~\ref{fig:stack}. In this case the reflection component
{\it decreases} for RX6 ($\chi^2/\nu$ = 17.42/6 for R fit to a constant
value), which makes more sense compared to the other values and is
consistent with a general trend that reflection decreases as the source
flux increases (Figure~\ref{fig:fvr}).  In other words, the reflected
flux appears to be consistently lagging behind the changes in the observed continuum.

\subsection{\SAX\ Results} \label{subsection:SAX_results}

\begin{figure}[p]
  \epsfxsize=3in
  \rotatebox{270}{\epsfbox[0 0 560 620]{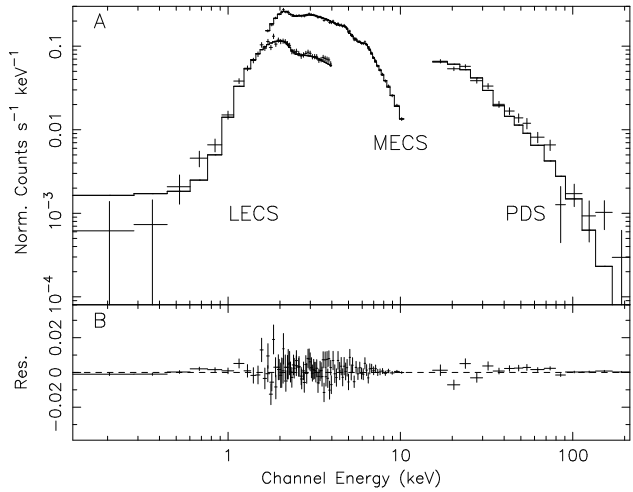}}
  \caption{\SAX\ data.  Panel A shows the data and best-fitting 
Compton reflection model, plus a Gaussian, folded through the 
instrumental response.  Panel B shows the residuals.}
  \label{fig:spec_SAX}
\end{figure}

The \SAX\ data are shown in Figure~\ref{fig:spec_SAX}. We use the same
model as for \RXTE\ except that the Fe K line width is allowed to vary
to account for any changes in the line profile. To account for the
relative normalization of the instruments, two additional free
parameters are added by allowing the MECS and PDS spectra to be offset
by a constant from the LECS spectrum. The total number of free
parameters is nine: $\Gamma$, $R$, A, iron line energy, iron line width
($\sigma$), iron line flux, \nH, the normalization of the MECS relative
to LECS ($N_{MECS}$), and the normalization of PDS relative to LECS
($N_{PDS}$).

The two best fits are shown in Table 3, the first with $\Gamma$ free and
the second with $\Gamma$ equal to $\Gamma_{(0.12-5.5)}$. The data do not
place a good constraint on the reflection fraction
(Figure~\ref{fig:fvr}), but there are notable changes in the spectrum
between the \SAX\ and \RXTE\ epochs. During the \SAX\ observation, the
source has fallen back to its \RXTE\ low-flux state, which is close to
the historical mean, and \nH\ has decreased by 40\%.  At the same time,
the Fe K line flux has fallen to almost one-half of its \RXTE\ value but
the equivalent width is similar to that seen with \RXTE. This result
implies that the features of X-ray reprocessing have tracked the
continuum on the longer timescale.

Other authors have commented on the fact that \SAX\ data poorly
constrain reflection (Wilkes, et al.\ 2001; Gliozzi, et al.\ 2001). 
Note that Perola et al. (2002) find similar \SAX\ error bars for
MCG$-$5-23-16.   We conclude that \RXTE\ provides a more sensitive 
measure of Compton reflection than \SAX.

\section{Discussion} \label{section:discussion}

Our \RXTE\ and \SAX\ observations, combined with data from the
literature, show that the X-ray spectrum of MCG$-$5-23-16 varies on time
scales from days to years.  The spectral changes are complex and, due to
the limited bandpass and resolution of \RXTE, the derived amounts of
absorption and reflection are coupled in a model-dependent fashion via
the photon index. From a model-independent comparison of spectra we
conclude that a $\sim20\%$ decrease in X-ray absorption is responsible
for a softening of the source in its highest flux state. Changes in
absorption on the order of 20\% to 80\% on timescales of less than 1
year are common in Seyfert 2 galaxies (Risaliti, Elvis \& Nicastro
2001). MCG$-$5-23-16 in particular has shown historical changes in \nH\
from $\sim 1.5 \times 10^{22}$ cm$^{-2}$ to $\sim 5 \times 10^{22}$
cm$^{-2}$. Taken together, the variability in \nH\ seen with \RXTE\ and
\SAX\ spans $1.5 \times 10^{22}$ cm$^{-2}$ to $3 \times 10^{22}$
cm$^{-2}$, which is consistent with its long-term behaviour (Risaliti,
Elvis \& Nicastro 2001).

Significant degeneracies between the photon index, the absorbing column
density and the amount of Compton reflection in standard reflection models  
can naively lead to false conclusions about spectral
variability.  These degeneracies occur as a result of the
way the three parameters trade against each other in the 
modeling process, especially for a source with large amounts of intrinsic
absorption.  We have attempted to deal with the MCG$-$5-23-16 spectra
in a way that allows us to eliminate the 
degeneracies as much as possble and focus on what we believe is the 
true variability in the source.   
We believe that our technique of examining the ratio of the high to
low-state spectra has isolated the model-dependent effects and 
has uncovered the true variability.

We find that the relative intensity of the Compton reflected continuum
to the directly-viewed continuum decreases as the source brightens on
timescales of days to months.  To avoid certain degeneracies in the
reflection model, we have assumed during the \RXTE\ epoch that $\Gamma$
is approximately constant. However, there has been much focus recently
on the intrinsic variability of the Seyfert continuum source.  In
particular, correlated changes between $\Gamma$, $R$ and the 2 to 10 keV
flux have been reported. Zdziarski, Lubi\'{n}ski \& Smith (1999) find
that Seyfert galaxies and X-ray binaries with
soft intrinsic spectra show much stronger reflection than do those with
hard spectra (although the authors above do not comment whether the flux
variability is self-consistent with the location of the proposed
reprocessor). Markowitz, Edelson \& Vaughan (2003) observe that the spectra
of Seyfert 1 galaxies tend to soften as the sources brighten.  
In addition, Gilfanov, Churazov \& Revnivtsev (1999) find
a strong correlation between $\Gamma$ and R for the galactic black hole
Cygnus X-1, and Lee et al. (2000) report that the flux and reflection
fraction track each other in the Seyfert 1 galaxy MCG $-$6-30-15. A
specific correlation between $\Gamma$ and $R$ can be explained 
theoretically by a
feedback between the X-ray emitting source and reflecting material. The
general idea is that the cold, neutral material (i.e. the reflecting
region) emits soft photons which irradiate the X-ray source and serve as
seed photons for Compton upscattering.

Because MCG$-$5-23-16 is fairly heavily absorbed, it is difficult to use
the \RXTE\ data to search for physical changes in $\Gamma$ and
correlations between $\Gamma$ and $R$. On the other hand, it is apparent
that $\Gamma$ does decrease between the \RXTE\ and \SAX\ epochs 1.5
years apart, which is consistent with the above correlation for 
Seyfert 1s since the
source flux also decreases between the two epochs.  However, during the
\RXTE\ epoch, any change in $\Gamma$ is {\it secondary} to the decrease
in \nH\ in terms of producing the observed spectral softening. Indeed,
not all studies of Seyfert galaxies show  a positive correlation between
flux, $\Gamma$ and $R$. Done, Madejski \& \.{Z}ycki (2000) find, using
\ASCA\ and \RXTE\ data, that the Seyfert 1 galaxy IC4329a shows a
marginal {\it anti-correlation} between $R$ and source flux. Examining
\RXTE\ data for Seyfert 2 galaxies, Georgantopoulos and Papadakis (2001)
find that although there is a correlation between $\Gamma$ and the
$2-10$ keV flux, there is a strong anti-correlation between $R$ and
flux.  Other work suggests similarly that reflection does not always
correlate with flux or photon index.  NGC 5548 has shown behavior where
$\Gamma$ is correlated with flux but not with $R$ (Chiang et al. 2000),
and NGC 5506 shows a tendency for $\Gamma$ to be larger for larger flux
states but a weak anticorrelation between flux and $R$ (Lamer, Uttley \&
McHardy 2000).

Reasons cited above for a lack of the expected correlation between $R$
and flux or $R$ and $\Gamma$ include the possibility that changes in
disk inner radial extent and/or ionization structure are small, or that
the variability is actually driven by changes in the seed photons that
are decoupled from the hard X-ray mechanism.  But in most cases the
authors assume that there must be an intrinsic correlation between the
state of the source (the ionizing spectrum) and the amount of
reprocessed emission.

A simple alternative explanation for a poor or 
inverse correlation between the intrinsic
flux and $R$ lies in the relative distance between the source and the
reprocessor.  First of all, if reprocessing occurs in a distant region
such as a molecular torus, then the reflected flux is controlled by the
time-averaged primary spectrum rather than the observed one (Malzac and
Petrucci 2001). Secondly, if there are multiple regions that contribute
reflected flux (such as a disk {\it and} a torus), then the reflected
spectrum will comprise a time-average of the portions that vary on short
timescales and the portions that vary on long timescales. Depending on
which portion dominates, the observed time lag will be weighted toward
that emission region.  Lamer, Uttley and McHardy (2000) have shown that
such a model explains well the lack of a correlation between flux and
$R$ in NGC 5506. By examining the relationship of the changes in the
spectral features of reprocessing (fluorescence and Compton reflection)
to changes in the source flux and continuum variability, it is possible
to infer limits on the size(s) of the region(s) that reprocess the
X-rays.

There is evidence for complicated X-ray reprocessing in MCG $-$5-23-16. 
The ASCA and XMM data show an Fe K$\alpha$ line that is triple-peaked with a
narrow component and broad, red and blue wings (Weaver et al.\ 1997;
Dewangan, Griffiths and Schurch 2003). The
broad wings are consistent with reprocessing due to an accretion disk
near the black hole while the line core is consistent with
reprocessing in the outer regions of the accretion disk or in the
molecular torus. Neither the \RXTE\ nor \SAX\ data presented here have
the resolution to distinguish these components of the iron line.
However, recent {\it Chandra} observations show that the Fe K line in
MCG$-$5-23-16 has a very narrow core with FWHM of $<$3,000 km s$^{-1}$
(Weaver 2001). The narrow Fe K line suggests a distant reprocessor.
There is also evidence for an obscuring torus at other wavelengths.
MCG$-$5-23-16 is heavily reddened in the optical with a weak, broad
component to H$\alpha$ (V\'{e}ron et al. 1980), and there is a broad
Pa$\beta$ line in the IR which is clear evidence for a broad line region
hidden by dust from our view at optical wavelengths (Goodrich, Veilleux,
\& Hill 1994).  If the narrow line arises in the torus, we expect some
of the Compton reflection to arise there as well.

We favor a scenario in which changes in $R$ are due to a time lag
following changes in the source flux; although, this picture is somewhat
complicated by noticeable changes in the absorbing column. We can place
limits on the light-crossing timescale; \RXTE\ observations imply a
lower limit of $\sim 1$ light day from the increase in $R$ when the
source flux drops to its lowest point between RX3 and RX4, while \SAX\
observations imply an upper limit of $\sim 1.5$ years from the fact that
the Fe K line has tracked the continuum. This places the bulk of the
reprocessing between $\sim 0.001$ pc (or $\sim 150 R_S$ for a $10^8
M_\sun$ black hole) and $\sim 0.55$ pc from the X-ray source.

\section{Conclusions} \label{section:conclusions}

We have examined spectral variability in MCG$-$5-23-16 based on \RXTE\
and \SAX\ observations. Based on the time delay between the continuum
and reprocessed emission, we derive limits on the response time between
changes in the 2 to 10 keV flux and reprocessed emission.  The bulk of
the reprocessed emission arises from a region that is located from 
between $\sim 1$ light day and $\sim 1.5$ light years away from the 
continuum source, which translates to a region that is between 0.001 pc
($150 R_s$ for a $10^8 M_\sun$ black hole) and 0.55 pc in size. The bulk of the
reprocessed emission must therefore arise in distant cold material,
perhaps in the outer regions of an accretion disk.

As part of our analysis, we demonstrate that the choice of the continuum 
photon index does influence the behavior 
of the Compton reflection fraction when
reflection models are applied without considering the broader complexity
of the source, such as the amount of underlying absorption.

We would like to thank Chris Reynolds for helpful discussions 
and the anonymous referee for providing many useful comments 
that improved the presentation of this paper.  This work was 
partly supported by NASA grant NAG5-4626.

%\begin{minipage}[t]{\textwidth}

\pagebreak

\begin{deluxetable}{crrrrrrrrr}
\tablecolumns{10}
\tabletypesize{\small}
\tablecaption{RXTE Spectral Fit Results: Fits for the 2 to 25 keV spectra with $\Gamma$ free}
\tablewidth{0pt}

\tablehead{
\colhead{} &
\colhead{} &
\colhead{} &
\colhead{} &
\colhead{} &
\colhead{Fe K$\alpha$} &
\colhead{Fe K$\alpha$} &
\colhead{Fe K$\alpha$} &
\colhead{} &
\colhead{} 
\\
\colhead{Spectrum\tablenotemark{a}} & 
\colhead{$n_H$\tablenotemark{b}} & 
\colhead{$\Gamma$\tablenotemark{c}} &
\colhead{R\tablenotemark{d}} &
\colhead{A\tablenotemark{e}} &
\colhead{Energy\tablenotemark{f}} &
\colhead{Flux\tablenotemark{g}} &
\colhead{W$_{K\alpha}$\tablenotemark{h}} &
\colhead{F$_{2-10}$\tablenotemark{i}} &
\colhead{$\chi^2_\nu$\tablenotemark{j}}
}
\startdata

RX1 &
$3.04_{-0.17}^{+0.17}$ &
$1.89_{-0.03}^{+0.03}$ &
$0.40_{-0.10}^{+0.11}$ &
$4.35_{-0.20}^{+0.22}$ &
$6.33_{-0.03}^{+0.05}$ &
$1.71_{-0.15}^{+0.19}$ &
$129_{-11}^{+14}$ &
$10.44_{-0.48}^{+0.51}$ &
1.32
\\
\\

RX2 &
$2.85_{-0.20}^{+0.19}$ &
$1.85_{-0.03}^{+0.03}$ &
$0.45_{-0.13}^{+0.13}$ &
$3.74_{-0.16}^{+0.21}$ &
$6.37_{-0.06}^{+0.06}$ &
$1.57_{-0.16}^{+0.20}$ &
$130_{-13}^{+17}$ &
$9.64_{-0.40}^{+0.54}$ &
1.41
\\
\\

RX3 &
$2.91_{-0.16}^{+0.15}$ &
$1.88_{-0.03}^{+0.03}$ &
$0.43_{-0.09}^{+0.10}$ &
$4.54_{-0.20}^{+0.13}$ &
$6.32_{-0.05}^{+0.05}$ &
$1.77_{-0.20}^{+0.14}$ &
$125_{-14}^{+10}$ &
$11.15_{-0.48}^{+0.32}$ &
1.39
\\
\\

RX4 &
$2.84_{-0.15}^{+0.15}$ &
$1.87_{-0.03}^{+0.03}$ &
$0.55_{-0.10}^{+0.11}$ &
$3.60_{-0.15}^{+0.16}$ &
$6.39_{-0.04}^{+0.04}$ &
$1.63_{-0.13}^{+0.13}$ &
$144_{-12}^{+11}$ &
$9.11_{-0.37}^{+0.40}$ &
1.11
\\
\\

RX5 &
$2.89_{-0.17}^{+0.17}$ &
$1.89_{-0.01}^{+0.03}$ &
$0.56_{-0.12}^{+0.13}$ &
$4.22_{-0.21}^{+0.12}$ &
$6.37_{-0.05}^{+0.06}$ &
$1.62_{-0.15}^{+0.19}$ &
$126_{-12}^{+15}$ &
$10.25_{-0.49}^{+0.30}$ &
1.32
\\
\\

RX6 &
$2.89_{-0.16}^{+0.16}$ &
$2.00_{-0.01}^{+0.02}$ &
$0.78_{-0.13}^{+0.15}$ &
$5.69_{-0.10}^{+0.30}$ &
$6.38_{-0.06}^{+0.06}$ &
$1.63_{-0.18}^{+0.19}$ &
$113_{-13}^{+13}$ &
$11.79_{-0.21}^{+0.62}$ &
0.89

\enddata

\tablenotetext{a}{The dates for the corresponding sprectra are: RX1, 24 April 1996; RX2, 28-29 July 1996; RX3, RX4, RX5, 27-30 November 1996; RX6, 10 January 1997.}
\tablenotetext{b}{Absorbing column density in units of $10^{22}$ cm$^{-2}$.}  
\tablenotetext{c}{Photon index of the intrinsic power-law spectrum. (f) indicates that this parameter was fixed in these fits.}  
\tablenotetext{d}{Fraction of reflected flux.}
\tablenotetext{e}{Power-law normalization in units of $10^{-2}$ photons keV$^{-1}$ cm $^{-2}$ s$^{-1}$ at 1 keV.}  
\tablenotetext{f}{Energy of the Fe K$\alpha$ line, in units of keV.}
\tablenotetext{g}{Flux of the Fe K$\alpha$ line in units of $10^{-4}$ photons cm$^{-2}$ s$^{-1}$.}  
\tablenotetext{h}{Equivalent width of Fe K$\alpha$ line, in units of eV.}
\tablenotetext{i}{2 to 10 keV flux in units of $10^{-11}$ ergs cm$^{-2}$ s$^{-1}$.}  
\tablenotetext{j}{Reduced $\chi^2$ for spectral fits, $\nu$ = number of degrees of freedom.}

\label{table:free_gamma}

\end{deluxetable}

\begin{deluxetable}{crrrrrrrrr}
\tablecolumns{10}
\tabletypesize{\small}
\tablecaption{RXTE Spectral Fit Results: Fits for the 2 to 25 keV spectra with $\Gamma$ fixed}
\tablewidth{0pt}

\tablehead{
\colhead{} &
\colhead{} &
\colhead{} &
\colhead{} &
\colhead{} &
\colhead{Fe K$\alpha$} &
\colhead{Fe K$\alpha$} &
\colhead{Fe K$\alpha$} &
\colhead{} &
\colhead{} 
\\
\colhead{Spectrum\tablenotemark{a}} & 
\colhead{$n_H$\tablenotemark{b}} & 
\colhead{$\Gamma$\tablenotemark{c}} &
\colhead{R\tablenotemark{d}} &
\colhead{A\tablenotemark{e}} &
\colhead{Energy\tablenotemark{f}} &
\colhead{Flux\tablenotemark{g}} &
\colhead{W$_{K\alpha}$\tablenotemark{h}} &
\colhead{F$_{2-10}$\tablenotemark{i}} &
\colhead{$\chi^2_\nu$\tablenotemark{j}}
}
\startdata

RX1 &
$2.99_{-0.10}^{+0.10}$ &
$1.88$ (f) &
$0.37_{-0.05}^{+0.05}$ &
$4.28_{-0.04}^{+0.04}$ &
$6.33_{-0.05}^{+0.05}$ &
$1.79_{-0.19}^{+0.12}$ &
$135_{-14}^{+9}$ &
$10.45_{-0.09}^{+0.09}$ &
1.30
\\
\\

RX2 &
$2.96_{-0.11}^{+0.11}$ &
$1.88$ (f) &
$0.53_{-0.06}^{+0.06}$ &
$3.89_{-0.04}^{+0.04}$ &
$6.38_{-0.06}^{+0.06}$ &
$1.55_{-0.18}^{+0.15}$ &
$128_{-15}^{+13}$ &
$9.62_{-0.10}^{+0.10}$ &
1.40
\\
\\

RX3 &
$2.91_{-0.09}^{+0.09}$ &
$1.88$ (f) &
$0.43_{-0.05}^{+0.04}$ &
$4.52_{-0.04}^{+0.04}$ &
$6.32_{-0.05}^{+0.05}$ &
$1.71_{-0.12}^{+0.19}$ &
$120_{-8}^{+14}$ &
$11.14_{-0.09}^{+0.09}$ &
1.36
\\
\\

RX4 &
$2.88_{-0.09}^{+0.09}$ &
$1.88$ (f) &
$0.57_{-0.05}^{+0.05}$ &
$3.64_{-0.03}^{+0.03}$ &
$6.39_{-0.04}^{+0.04}$ &
$1.64_{-0.16}^{+0.09}$ &
$145_{-14}^{+8}$ &
$9.11_{-0.08}^{+0.08}$ &
1.09
\\
\\

RX5 &
$2.82_{-0.10}^{+0.10}$ &
$1.88$ (f) &
$0.50_{-0.05}^{+0.05}$ &
$4.11_{-0.02}^{+0.04}$ &
$6.36_{-0.05}^{+0.05}$ &
$1.63_{-0.12}^{+0.19}$ &
$128_{-10}^{+15}$ &
$10.26_{-0.05}^{+0.09}$ &
1.31
\\
\\

RX6 &
$2.32_{-0.10}^{+0.09}$ &
$1.88$ (f) &
$0.35_{-0.05}^{+0.05}$ &
$4.65_{-0.04}^{+0.04}$ &
$6.32_{-0.05}^{+0.05}$ &
$1.99_{-0.23}^{+0.13}$ &
$139_{-16}^{+9}$ &
$11.90_{-0.10}^{+0.11}$ &
1.80

\enddata

\tablenotetext{a}{The dates for the corresponding spectra are: RX1, 24 April 1996; RX2, 28-29 July 1996; RX3, RX4, RX5, 27-30 November 1996; RX6, 10 January 1997.}
\tablenotetext{b}{Absorbing column density in units of $10^{22}$ cm$^{-2}$.}  
\tablenotetext{c}{Photon index of the intrinsic power-law spectrum. (f) indicates that this parameter was fixed in these fits.}  
\tablenotetext{d}{Fraction of reflected flux.}
\tablenotetext{e}{Power-law normalization in units of $10^{-2}$ photons keV$^{-1}$ cm $^{-2}$ s$^{-1}$ at 1 keV.}  
\tablenotetext{f}{Energy of the Fe K$\alpha$ line, in units of keV.}
\tablenotetext{g}{Flux of the Fe K$\alpha$ line in units of $10^{-4}$ photons cm$^{-2}$ s$^{-1}$.}  
\tablenotetext{h}{Equivalent width of Fe K$\alpha$ line, in units of eV.}
\tablenotetext{i}{2 to 10 keV flux in units of $10^{-11}$ ergs cm$^{-2}$ s$^{-1}$.}  
\tablenotetext{j}{Reduced $\chi^2$ for spectral fits, $\nu$ = number of degrees of freedom.}

\label{table:fixed_gamma}

\end{deluxetable}

\begin{deluxetable}{crrrrrrrrrr}
\tablecolumns{11}
\tablewidth{0pt}
\tabletypesize{\small}
\tablecaption{SAX Spectral Fit Results \label{table:sax}}

\tablehead{
\colhead{} &
\colhead{} &
\colhead{} &
\colhead{} &
\colhead{} &
\colhead{Fe K$\alpha$} &
\colhead{Fe K$\alpha$} &
\colhead{Fe K$\alpha$} &
\colhead{Fe K$\alpha$} &
\colhead{} &
\colhead{}
\\
\colhead{Fit} &
\colhead{$n_H$\tablenotemark{a}} & 
\colhead{$\Gamma$\tablenotemark{b}} &
\colhead{R\tablenotemark{c}} &
\colhead{A\tablenotemark{d}} &
\colhead{Energy\tablenotemark{e}} &
\colhead{$\sigma$\tablenotemark{f}} &
\colhead{Flux\tablenotemark{g}} &
\colhead{W$_{K\alpha}$\tablenotemark{h}} &
\colhead{F$_{2-10}$\tablenotemark{i}} &
\colhead{$\chi^2_\nu$\tablenotemark{j}} 
}
\startdata

1\tablenotemark{k} 		&
1.66$^{+0.06}_{-0.06}$ 	& 
1.73$^{+0.02}_{-0.04}$ 	& 
0.22$^{+0.37}_{-0.22\tablenotemark{k}}$ & 
2.05$^{+0.12}_{-0.10}$ 	& 
6.40$^{+0.08}_{-0.08}$ 	& 
0.19$^{+0.14}_{-0.19\tablenotemark{l}}$ & 
1.02$^{+0.30}_{-0.30}$	& 
128$^{+38}_{-38}$  		& 
9.15$^{+0.52}_{-0.45}$ 	& 
1.01
\\
\\			
2\tablenotemark{m} 		&		
1.70$^{+0.04}_{-0.05}$ 	& 
1.77 (f)               	& 
0.48$^{+0.22}_{-0.21}$ 	& 
2.12$^{+0.04}_{-0.04}$ 	& 
6.40$^{+0.08}_{-0.08}$ 	& 
0.18$^{+0.15}_{-0.18\tablenotemark{l}}$ & 
0.86$^{+0.45}_{-0.22}$	& 
107$^{+55}_{-28}$  		& 
9.13$^{+0.16}_{-0.18}$ 	& 
1.01 

\enddata

\tablenotetext{a}{Absorbing column density in units of $10^{22}$ cm$^{-2}$.}  
\tablenotetext{b}{Photon index of the intrinsic power-law spectrum. (f) indicates that this parameter was fixed in these fits.}  
\tablenotetext{c}{Fraction of reflected flux.}
\tablenotetext{d}{Power-law normalization in units of $10^{-2}$ photons keV$^{-1}$ cm $^{-2}$ s$^{-1}$ at 1 keV.}  
\tablenotetext{e}{Energy of the Fe K$\alpha$ line, in units of keV.}
\tablenotetext{f}{Physical width of the Fe K$_\alpha$ line, in units of eV.}
\tablenotetext{g}{Flux of the Fe K$\alpha$ line in units of $10^{-4}$ photons cm$^{-2}$ s$^{-1}$.}  
\tablenotetext{h}{Equivalent width of Fe K$\alpha$ line, in units of eV.}
\tablenotetext{i}{2 to 10 keV flux from MECS in units of $10^{-11}$ ergs cm$^{-2}$ s$^{-1}$.} 
\tablenotetext{j}{Reduced $\chi^2$ for spectral fits, $\nu$ = number of degrees of freedom.}
\tablenotetext{k}{All parameters are free for this fit.  The normalization of the MECS spectra relative to the LECS is $N_{MECS} = 1.339^{+0.029}_{-0.028}$, and the normalization of the PDS spectra relative to the LECS is $N_{PDS} = 1.286^{+0.170}_{-0.194}$}
\tablenotetext{l}{A hard limit in the fit parameter was reached before a $\delta \chi^2 = 2.706$.}
\tablenotetext{m}{For this fit, $\Gamma$ is fixed at its 2-5.5 keV value and all other parameters are free.  The normalization of the MECS spectra relative to the LECS is $N_{MECS} = 1.338^{+0.031}_{-0.030}$, and the normalization of the PDS spectra relative to the LECS is $N_{PDS} = 1.184^{+0.153}_{-0.129}$}

\end{deluxetable}

\end{document}